\documentclass[12pt,a4paper]{article}
\usepackage[english]{babel}
\usepackage[utf8]{inputenc}
\usepackage{amsmath}
\usepackage{amsfonts}
\usepackage{amssymb}
\usepackage{authblk}
\usepackage{graphicx}
\usepackage{relsize}
\usepackage{hyperref}
\usepackage{lineno}
\usepackage{color}
\usepackage[dvipsnames]{xcolor}

\newcommand{\beq}{\begin{equation}}
\newcommand{\beqn}{\begin{eqnarray}}
\newcommand{\eeq}{\end{equation}}
\newcommand{\eeqa}{\end{eqnarray}}
\newcommand{\nn}{\nonumber}

\newcommand{\pgb}{\mathrel{\{\![}}
\newcommand{\pbg}{\mathrel{]\!\}}}

\newcommand{\inter}{\mathrel{\mathlarger{\mathlarger{\mathlarger{\lrcorner}}}}}

\newcommand{\mfc}{\overset{\,p}{X}}

\newcommand{\pform}{\overset{\,p}{F}}
\newcommand{\qform}{\overset{\,q}{F}}

\newcommand{\ene}{\overset{\,n-p}{F}}

\newcommand{\npfield}{\overset{\,n-p}{X}}
\newcommand{\nqfield}{\overset{\,n-q}{X}}

\title{Polysymplectic formulation for topologically massive Yang-Mills field theory}

\author[1]{Jasel Berra-Montiel\thanks{jberra@fc.uaslp.mx}}
\author[1]{Eslava del R\'io\thanks{eslava@fc.uaslp.mx}}
\author[1,2]{Alberto Molgado\thanks{molgado@fc.uaslp.mx}}

\affil[1]{
Facultad de Ciencias, Universidad Autonoma de San Luis Potosi,
Av. Salvador Nava S/N Zona Universitaria,
San Luis Potosi, SLP, 78290 Mexico
}
\affil[2]{
Dual CP Institute of High Energy Physics, Mexico  
}

\begin{document}

%\linenumbers

%%%%%%%%%%%%%%%%%%%%% Publisher's Area please ignore %%%%%%%%%%%%%%%
%
%\catchline{}{}{}{}{}
%
%%%%%%%%%%%%%%%%%%%%%%%%%%%%%%%%%%%%%%%%%%%%%%%%%%%%%%%%%%%%%%%%%%%%

\maketitle

\begin{abstract}
We analyze the De Donder-Weyl covariant field equations for the
topologically massive Yang-Mills theory. 
These equations are obtained through the Poisson-Gerstenhaber bracket 
described within the polysymplectic 
framework. Even though the Lagrangian defining 
the system of our interest is singular, we
show that by appropriately choosing the polymomenta one may obtain an equivalent regular Lagrangian, thus avoiding the standard analysis of constraints.  Further, our simple treatment 
allows us to only consider the privileged $(n-1)$-forms
in order to obtain the correct field equations, in opposition to certain examples found in the literature.
\end{abstract}

%\ccode{Keywords: Gauge theory, Yang-Mills, Chern-Simons,  Multisymplectic formulation, De Donder-Weyl Hamiltonian, Poisson-Gerstenhaber bracket}

%\ccode{PACS numbers: 11.15.Kc, 03.50.Kk, 12.10.-g, 11.10.Ef}

%\ccode{AMS classification scheme numbers: 70S15, 70S05}

\section{Introduction}

Non-Abelian gauge models like Yang-Mills and 
Chern-Simons field theories are very well suited to describe 
a huge diversity of phenomena in physics. On the one hand, the Yang-Mills field theory has been completely successful describing the standard model of particle physics~\cite{ryder},  while the non-Abelian Chern-Simons field theory, on the other hand, is commonly associated to a wide variety of areas ranging from condensed matter and solid state physics 
to general relativity~\cite{witten1,witten2,hag,cs-jackiw,jackiw,banados1,banados2}.
The topologically massive Yang-Mills field 
theory is obtained by adding a Chern-Simons invariant to the Yang-Mills theory~\cite{deser1,deser2,deser3}, thus resulting 
in a mechanism to generate gauge field mass.
As it is well known, topologically massive Yang-Mills theory provides a link between field theories
and knot theory, and
also has been introduced to model 
specific issues in several areas of physics.  
For example, it has been related to the 
quantum Hall effect~\cite{qhe} and 
superconductivity~\cite{wil}, while in the 
gravitational context is related to the 
topological massive gravity~\cite{deguchi,gr-deser,yildrim}.

At the classical level, equations of motion for the topologically massive Yang-Mills theory has 
been studied from different perspectives, including the standard 
Dirac-Hamiltonian~\cite{evens} and the 
Hamilton-Jacobi~\cite{tmym}
approaches. In this article we analyze
this topologically massive theory from the modern viewpoint 
of the polysymplectic framework 
for field theories 
(see~\cite{Gotay,helein,for,kijo,cari,kan,roman} for generalities and the geometric 
description, and~\cite{kan,vey,marsdena,helein1} for some specific 
examples). Even though some of the 
various versions of the multisymplectic formalism\footnote{In the literature, the term multisymplectic is commonly associated to generic structures on fiber bundles which 
do not distinguish vertical and horizontal subspaces.  On the contrary, the term polysymplectic is used 
whenever an explicit decomposition into these subspaces is considered.}  may differ in their geometric 
ingredients,
the vast majority of state-of-the-art versions
rely on its construction on a jet bundle structure which, being 
naturally covariant, offers an outstanding  
perspective for the analysis of physically motivated field theories.  Also, one of the main features of the jet bundle structure is that 
a field theory may be seen as a finite dimensional extended model. 
Besides, the polysymplectic 
formalism  is strongly based on the De Donder-Weyl equations which are covariant in the sense that there is no privileged direction in the definition of the polymomenta (as opposed to the standard symplectic Hamiltonian approach where the time variable is treated as a privileged coordinate in the momenta definition). 
The De Donder-Weyl equations may be 
obtained from a a generalization of the standard Poisson bracket, known as the Poisson-Gerstenhaber bracket and defined for Hamiltonian forms under the co-exterior product. Indeed, the 
Hamiltonian $(n-1)$-forms play a special role under this Poisson-Gerstenhaber bracket as they
guarantee closure  of the bracket under the co-exterior product defined on a $(n-1)$-dimensional subspace and they also are necessary in order to get the correct 
equations of motion as described in~\cite{kan,kan1,kan2,kan3,helein2} (see also~\cite{zapata} for a discussion within 
the multisymplectic formalism for lattice field theories).

In order to proceed our analysis for the topologically massive Yang-Mills theory 
we start by testing first the polysymplectic
formalism for the Yang-Mills and the 
Chern-Simons theories, and then we 
continue systematically.   We thus find that, 
by judiciously decomposing the polymomenta into its symmetric and anti-symmetric parts,
the correct equations of motion may be obtained
by considering only the Hamiltonian $(n-1)$-forms.  Further, for the three cases we obtained 
a vanishing divergence of the symmetric parts 
of the polymomenta.  This in turn allow us to introduce 
an appropriate transformation of the polymomenta  inducing an equivalent De Donder-Weyl Hamiltonian
from which the equations of motion for each field may be straightforwardly obtained.  
Our treatment may be confronted 
with the analysis of the Maxwell field in 
reference~\cite{kan} where, adversely, the  Hamiltonian $(n-1)$-forms are not considered as 
the fundamental forms in order to obtain the 
correct equations of motion.

The rest of the article is organized as follows. In Section~\ref{sec:multi} we summarize the polysymplectic  
formalism based on the explicit decomposition
of the fiber bundles into its vertical 
and horizontal parts
and 
describe the De Donder-Weyl equations of motion in order to briefly introduce all the required background and to set the our notation. In Section~\ref{sec:nonabelian} we apply the polysymplectic formalism to the Yang-Mills, Chern-Simons and topologically massive Yang-Mills field theories.  We emphasize that the 
three non-Abelian field theories contain analogous structures
which allow us to describe these models systematically. Finally, in Section~\ref{conclusion} we include some concluding remarks.

\section{Polysymplectic  structure and the Poisson-Gerstenhaber bracket}
\label{sec:multi}

In this Section we briefly introduce  the 
De Donder-Weyl Hamiltonian from the perspective of the 
polysymplectic formalism and its relation to the Poisson-Gerstenhaber bracket.  By considering simplicity as 
our guiding principle, we proceed as close as possible to references~\cite{kan,kan1,kan2,kan3}.  We encourage the reader to check these references 
for further details and, also, reference~\cite{paufler}
for a technical construction of the 
vertical-horizontal splitting in the context of the polysymplectic formalism  for nontrivial fiber bundles.

To start, we will consider an arbitrary smooth $n$-dimensional spacetime manifold $\mathcal{M}$ with 
local coordinates $\{x^\mu\},\ \mu=1,\ldots n$, and volume form $\omega$.  We also will consider the fibered  manifold
$(\mathcal{E},\pi, \mathcal{M})$, where $\mathcal{E}$ denotes the total space manifold with local coordinates $\{\phi^a\}$  (local sections around $p\in\mathcal{M}$) which may 
be physically interpreted as the 
classical gauge fields associated to 
a given theory ($a=1,\ldots,m$ 
denoting the set of internal degrees of freedom), and the map $\pi:\mathcal{E}\rightarrow\mathcal{M}$ stands for the 
canonical projection.  Finally, we will consider the 
first jet manifold of $\pi$, $J^1\mathcal{E}$, with local coordinates 
$(x^\mu,\phi^a,\phi^a_\mu)$, where $\phi^a_\mu:=\partial\phi^a/\partial x^\mu$
stand for the field derivative coordinates. In what follows, we will
identify the first jet manifold with the 
configuration space of the theory.

In order to describe the dynamics of a given theory we will consider the Lagrangian density 
$L : J^1\mathcal{E}\rightarrow  \mathbb{R}$ 
as the smooth function of the configuration space such
that $L(j_p^1 \phi) =  
L(\phi^a, \partial_\mu \phi^a, x^\mu)$, where 
$j_p^1$ is the first prolongation of the jet bundle. Next, we introduce the polymomenta given by
 \begin{equation} \label{poly}
\pi^\mu_a :=  \frac{\partial L}{\partial(\partial_\mu \phi^a)} \,.
\end{equation}
The polymomentum phase space is endowed with a canonical $n$-form $\Theta_{\mathrm{DW}}$ known as the Poincaré-Cartan form, given by \cite{katsup,Hubert,Gotay1}
\begin{equation}
\Theta_{\mathrm{DW}} = \pi^\mu_a d\phi^a\wedge\omega_\mu - H_{\mathrm{DW}}\omega \,,
\end{equation} 
where $\omega_\mu := \partial_\mu\inter\omega$ is the basis for the $(n-1)$-form subspace, and
the De Donder-Weyl Hamiltonian function 
$H_{\mathrm{DW}}$
is obtained by means of the covariant Legendre transformation
\begin{equation}\label{legen}
H_{\mathrm{DW}}(\phi^a, \pi^\mu_a, x^\mu) := \pi^\mu_a \partial_\mu \phi^a - L \,.
\end{equation}
Calculating the exterior differential of the Poincar\'e-Cartan form results in the $(n+1)$-form
\begin{equation}
\Omega_{\mathrm{DW}} := 
d \Theta_{\mathrm{DW}}
= d\pi^\mu_a\wedge d\phi^a\wedge\omega_\mu - dH_{\mathrm{DW}}\wedge\omega \,.
\end{equation}
As described in references~\cite{kijo,cari,lie,sar}, the classical dynamics of the fields is essentially encoded in the vertical components of the multivector field annihilating the canonical $(n+1)$-form $\Omega_{\mathrm{DW}}$. Thus, we will only consider the vertical part of the Poincar\'e-Cartan $n$-form
\begin{equation}
\Theta^V_{\mathrm{DW}} := \pi^\mu_a d\phi^a\wedge\omega_\mu \,.
\end{equation}
Now, let us define the vertical exterior differential as 
\beqn
d^V\Phi =\frac{1}{p!}\partial_v \Phi^{M_1\dots M_p}dz^v\wedge dz^{M_1}\wedge\dots\wedge dz^{M_p} \,,
\eeqa
where $z^M:=(z^v,x^\mu)=(\phi^a, \pi^\mu_a, x^\mu)$ stands for the first jet bundle local coordinates. 
In this way we are able to calculate the vertical exterior differential of $\Theta^V_{\mathrm{DW}}$ 
\begin{equation}\label{polyform}
\Omega^V_{\mathrm{DW}} = d^V \Theta^V_{\mathrm{DW}} = -d\phi^a\wedge d\pi^\mu_a\wedge\omega_\mu \,.
\end{equation}
This $\Omega^V_{\mathrm{DW}}$ is simply known as the 
polysymplectic $(n+1)$-form.

Let $\mfc~\!\!^{V}$ be a vertical multivector field, that is, a $p$-multivector field such that it has one vertical and $(p-1)$ horizontal indices, namely
$\mfc~\!\!^{V}=\mfc~\!\!^{v\mu_1\ldots\mu_{p-1}}(z^M)\partial_v\wedge\partial_{\mu_1}\wedge\cdots\wedge\partial_{\mu_{p-1}}$. We will call the vertical multivector field $\mfc~\!\!^{V}$ a
Hamiltonian multivector if there exists an horizontal $(n-p)$-form $\ene$ such that
\begin{equation}\label{hamf}
\mfc~\!\!^{V}\inter\Omega^{V}_{\mathrm{DW}} = d^V \ene \,. 
\end{equation} 
From now on, we will assume that all the multivector fields are vertical. As discussed in~\cite{kan,kan1,kan2,kan3}, the space of Hamiltonian forms results not closed with respect to the exterior product, but it is closed with respect to the co-exterior product defined by
\begin{equation}\label{pgproduct}
\pform\bullet\qform := \ast^{-1}(\ast\pform\wedge\ast\qform) \,,
\end{equation}
where $\ast$ stands for the standard Hodge operator.
In addition, for a pair of Hamiltonian forms one may define the Poisson-Gerstenhaber bracket, $\pgb \mathlarger{\mathlarger{\mathlarger{\cdot}}} , \mathlarger{\mathlarger{\mathlarger{\cdot}}} \pbg$, 
\begin{equation}\label{pgbracket}
\pgb \pform_1, \qform_2 \pbg = (-1)^{n-p} \npfield_1\inter\nqfield_2\inter\Omega^V_{\mathrm{DW}} \,.
\end{equation}
One may  note that the Hamiltonian forms provided with the Poisson-Gerstenhaber bracket, $\pgb\cdot\,,\cdot\pbg$, 
close as a graded Poisson-Gerstenhaber algebra
under the co-exterior product defined in~(\ref{pgproduct}).
In addition, we also note from the bracket~(\ref{pgbracket}) that the  Hamiltonian 
$(n-1)$-forms will play a primordial role within this polysymplectic formalism as for them the bracket 
clearly results closed.  This relevance is associated 
to the fact that we may construct Noether currents 
by means of the $(n-1)$-forms in order to 
construct the observables for a given theory~\cite{helein2,zapata}. 
We may use the Poisson-Gerstenhaber bracket to induce the standard relations among the canonically conjugate variables
 \begin{equation}\label{cre}
 \pgb\pi^\mu_a\omega_\mu, \phi^b\pbg = \delta^b_a, \hspace{4mm} \pgb\pi^\mu_a\omega_\mu, \phi^b\omega_\nu\pbg = \delta^b_a\omega_\nu, \hspace{4mm} \pgb\pi^\mu_a, \phi^b\omega_\nu\pbg = \delta^b_a\delta^\mu_\nu \,.
 \end{equation}  
%\begin{eqnarray}
%\pgb\pform, \qform\pbg  &=& -(-1)^{g_1 g_2} \pgb\qform, \pform\pbg \,, \nonumber \\
%(-1)^{g_1 g_3}\pgb\pform, \pgb\qform, \rform\pbg\pbg &=& (-1)^{g_1 g_2}\pgb\qform, \pgb\rform, \pform\pbg\pbg % 
% + (-1)^{g_2 g_3}\pgb\rform, \pgb\pform, \qform\pbg\pbg \,, \nonumber \\
% \pgb \pform, \qform\bullet\rform\pbg &=& \pgb\pform, \qform\pbg\bullet\rform + (-1)^{g_1(g_2+1)}\qform\bullet\pgb\pform, \rform\pbg \,, 
%\end{eqnarray}
%where $g_1= n-p-1$, $g_2 = n-q-1$ and $g_3 = n-r-1$.\\
From our particular point of view, however, a very important application of the Poisson-Gerstenhaber bracket is that it allows us to write the De Donder-Weyl field equations for an arbitrary  Hamiltonian
$(n-1)$-form $F = F^\mu\omega_\mu$ by means of the relation
\begin{equation}\label{dwfe}
\textbf{d}\bullet F = -\sigma(-1)^n \pgb H_{\mathrm{DW}}, F\pbg  + d^h \bullet F \,,
\end{equation}
where $\sigma = \pm 1$.~\footnote{The plus and the minus signs
stand for Euclidean and Minkowskian signatures of the 
spacetime manifold $\mathcal{M}$, respectively.}  The operation $\textbf{d}\bullet$ is known as the total co-exterior differential and for an arbitrary $p$-form $\pform$ reads
\begin{equation}
\textbf{d}\bullet\pform := \frac{1}{(n-p)!}\partial_v F^{\mu_1\dots\mu_{n-p}}\partial_\mu z^v dx^\mu\bullet\partial_{\mu_1\dots\mu_{n-p}}\inter\omega + d^h\bullet \pform \,,
\end{equation}
and $d^h \bullet$ denotes the horizontal co-exterior differential given by
\begin{equation}
d^h\bullet\pform := \frac{1}{(n-p)!}\partial_\mu F^{\mu_1\dots\mu_{n-p}} dx^\mu\bullet\partial_{\mu_1\dots\mu_{n-p}}\inter\omega \,. 
\end{equation}

Finally, by considering the canonical brackets~(\ref{cre}) and the total co-exterior differential~(\ref{dwfe}) for the 
components of the canonical variables, $\pi^\mu_a$ and $\phi^a$, we obtain
the De Donder-Weyl field equations
\begin{equation}\label{dweqs}
\partial_\mu \pi^\mu_a = -\frac{\partial H_{\mathrm{DW}}}{\partial \phi^a} \,, \hspace{15mm}\partial_\mu \phi^a = \frac{\partial H_{\mathrm{DW}}}{\partial \pi^\mu_a} \,.
\end{equation}
It is straightforward to show that the relations \eqref{dweqs} are equivalent to the Lagrangian field equations if $L$ is hyperregular, that is, 
whenever the covariant Legendre transformation
is a diffeomorphism~\cite{Gotay,katsup,HJT,GOCF}, 
thus, hyperregular Lagrangians on the first jet manifold, $J^1\mathcal{E}$, induce well defined De Donder-Weyl Hamiltonian systems on
the dual jet $J^{1*} \mathcal{E}$, and vice versa.
Even though for the cases we are interested below
this last condition does not follow, it is possible 
to extend this formalism to these kind of systems due
to the inherent symmetries, as will be discussed.

\section{Non-Abelian field theories}
\label{sec:nonabelian}

The main purposes of this section are to describe, 
within the polysymplectic framework, the field equations associated to the topologically massive Yang-Mills field theory, on the one side, and to 
analyze the emerging constraints for this model, on the other side. In order to achieve this, first we will 
develop separately the polysymplectic and Poisson-Gerstenhaber structures, as described in the previous section, for a pair of related models, namely the Yang-Mills field theory and the non-Abelian Chern-Simons  topological field theory. After doing that, we will couple the Yang-Mills and the Chern-Simons theories to obtain the  model of our interest. From now on, we will assume that all the fields of our interest are
described in a three dimensional background space-time manifold $\mathcal{M}$ endowed with a Minkowski metric $\eta_{\mu\nu} =$ diag(1, -1, -1) and Latin indices which describe internal degrees of freedom are raised and lowered by the corresponding metric in the internal space. 

\subsection{Yang-Mills field theory}

The Yang-Mills field theory has been previously analyzed
within the context of the multisymplectic formulation in references~\cite{YM1},~\cite{YM2}, where the main discussion was guided towards the quantum representation, and the 
constrained structure for this model was only considered heuristically.  From our point of view, our intention is to elucidate the 
role of the constraint related to the
symmetric part of polymomenta explicitly.  To this end, 
let us begin with the well known $2+1$ dimensional Yang-Mills model which is defined by means of the 
Lagrangian~\cite{sunder,kosyakov}
\begin{equation}
L_{\mathrm{YM}} = -\frac{1}{4} F_{\mu\nu}^a F^{\mu\nu}_a \,, 
\end{equation}
where  $\mu = 1, 2, 3$ denote space-time indices while  lower-case Latin indices denote  internal degrees of freedom. The components of the field strength are given by 
\begin{equation}\label{fcomp}
F_{\ \mu\nu}^a = \partial_\mu A_\nu^a - \partial_\nu A_\mu^a + g f^a_{\ bc} A_\mu^b A_\nu^c \,.
\end{equation}
Here $A^a_\nu$ are the gauge or Yang-Mills fields, 
and generalize the vector potential in electrodynamics,  $g$ is a coupling constant and $f_{abc}$ are 
the structure constants of the Lie 
algebra associated to the internal symmetry group. 
Now, 
by adapting definition \eqref{poly} to our case, we obtain the polymomenta
\begin{equation}\label{ygpoly}
\pi^{\mu\nu}_a = \frac{\partial L_{\mathrm{YM}}}{\partial(\partial_\mu A^a_\nu)}= -F^{\mu\nu}_a \,,
\end{equation}
which, due to the anti-symmetry of the field strength $F^{\mu\nu}_a$,  yields the conditions
\begin{equation}\label{pcym}
\pi^{(\mu\nu)}_a \approx 0 \,.
\end{equation}
(Note that these conditions result analogous to the 
primary constraints within the Dirac 
formalism~\cite{henneaux}.  
Henceforth, and by a slight abuse of language, we will simply refer to conditions~\eqref{pcym} as primary constraints, and 
will endow the weak equality symbol with the same meaning as 
in Dirac approach.  See~\cite{kan3} for further details in the Dirac treatment for constraints within the polysymplectic formalism).
In the following, we will consider the symmetric and anti-symmetric parts of our polymomentum and field variables, so we can get the appropriate field equations.
By means of the Legendre transformation \eqref{legen} we obtain the De Donder-Weyl Hamiltonian for the Yang-Mills theory
\beqn
H_{\mathrm{DW}}^{\mathrm{YM}}(A, \pi, x) 
& = & 
\pi_a^{[\mu\nu]} \partial_{[\mu}
A^a_{\nu]} - L_{\mathrm{YM}}
\nn \\
& = & -\frac{1}{4} \pi_{[\mu\nu]}^a \pi^{[\mu\nu]}_a -\frac{g}{2} f^{\ bc}_a A^\mu_b A^\nu_c \pi_{[\mu\nu]}^a \,.
\eeqa
In analogy to the standard Dirac formalism for singular Lagrangians~\cite{henneaux}, we define the total De Donder-Weyl Hamiltonian, $\widetilde{H}_{\mathrm{DW}}^{\mathrm{YM}}$, as the De Donder-Weyl Hamiltonian subject to 
the primary constraints~\eqref{pcym}, that is, 
 \begin{equation}\label{ymth}
 \widetilde{H}_{\mathrm{DW}}^{\mathrm{YM}} = 
H_{\mathrm{DW}}^{\mathrm{YM}} + \lambda^a_{\mu\nu}\pi^{(\mu\nu)}_a \,,
 \end{equation}
 where $\lambda^a_{\mu\nu}$ are Lagrange multipliers
 enforcing the constraints~\eqref{pcym}. Of course, the Lagrange multipliers $\lambda^a_{\mu\nu}$ are symmetric in Greek 
 indices.  
We will use this total De Donder-Weyl Hamiltonian in order to analyze the 
correct equations of motion  for the system.  
Indeed, by considering the $(n-1)$-forms 
$A^{a\nu}_\mu := A^a_\mu\omega^\nu$ and 
$\pi^\nu_a := \pi^{\mu \nu}_a\omega_\mu $  as the canonically conjugate variables (see~\eqref{cre}), the field equations~\eqref{dwfe} read 
\beqn
\textbf{d}\bullet A^{\mu\nu}_a 
& = & 
- \pgb \widetilde{H}_{\mathrm{DW}}^{\mathrm{YM}}, A^{\mu\nu}_a\pbg  %+ d^h \bullet A^{\mu\nu}_a  
\nn\\
& = &
-\frac{1}{2} \pi^{[\mu\nu]}_a -\frac{g}{2} f^{bc}_a A^\mu_b A^\nu_c \,,  
\nn\\
\textbf{d}\bullet \pi_a^\mu 
& = & 
- \pgb \widetilde{H}_{\mathrm{DW}}^{\mathrm{YM}}, \pi_a^\mu\pbg  %+ d^h \bullet \pi_a^\mu  
\nn\\
& = &
 -gf^c_{ab} A^b_\nu \pi^{[\nu\mu]}_c \,,
\eeqa  
where the horizontal co-exterior differential terms
in~(\ref{dwfe}) automatically vanish as the background spacetime we are considering is Minkowski.  In the subsequent models, this will also be the case.
By considering the symmetric and anti-symmetric parts of these equations of motion we straightforwardly obtain the set of relations
\begin{eqnarray}
\label{eq:ym-identities}
\partial^{[\mu} A^{\nu]}_a 
& = & 
-\frac{1}{2} \pi^{[\mu\nu]}_a -\frac{g}{2} f^{bc}_a A^\mu_b A^\nu_c \,,
\nn\\
\partial^{(\mu} A^{\nu)}_a 
& = & 
\lambda_a^{(\mu\nu)} \,, 
\nn\\
\partial_\nu \pi^{[\nu\mu]}_a  -gf^c_{\ ab} A^b_\nu \pi^{[\nu\mu]}_c 
& = & 
0 \,,
\nn\\ 
\partial_\nu \pi^{(\nu\mu)}_a 
& = & 0 \,.
\end{eqnarray}  
These relations allow us to fully characterize the dynamics of the system.  
Certainly, the first line of~\eqref{eq:ym-identities} only stands for the definition 
of the polymomenta~\eqref{ygpoly}.  The second line fixes  the Lagrangian multipliers in terms of the symmetric derivatives of the 
gauge field.   By defining the 
covariant derivative for the gauge group as
\beqn
\label{eq:covderivative}
D^{b}_{\mu a} :=\delta^{b}_a\partial_\mu - gf^{b}_{\ ac} A^c_\mu \,, 
\eeqa
where the $\delta$ stands for the identity for the generators of the internal group, and by considering relation~\eqref{ygpoly}, $\pi^{[\mu\nu]}_a = -F^{\mu\nu}_a$, we may identify
the third line above as the standard 
Yang-Mills field 
equations for the field strength $F^{\nu\mu}_a$, that is, 
\beq
(D_\nu F^{\nu\mu})_a = 0  \,.
\eeq
Finally, the last line may be interpreted, 
together with constraints~\eqref{pcym}, as
the fact that the symmetric part of the polymomentum does not contain information about the dynamics of the model.

We observe that in the De Donder-Weyl formalism, the variation of the Hamiltonian only determines the divergence of the polymomenta (\ref{dweqs}), therefore, the equations of motion given in (\ref{eq:ym-identities}), remain invariant if the polymomenta transform as 
\begin{equation}
\pi^{\mu\nu}_a \mapsto\pi'^{\mu\nu}_{a}:= \pi^{\mu\nu}_{a}-\xi^{\mu\nu}_{a},
\end{equation}
where the term $\xi^{\mu\nu}_{a}$ must satisfies the divergenceless condition, $\partial_{\mu}\xi^{\mu\nu}_{a}=0$. In order to see this, let us define a new Lagrangian, $L'_{\mathrm{YM}}$, given by
\begin{equation}
L'_{\mathrm{YM}}=L_{\mathrm{YM}}-\xi^{\mu\nu}_{a}\partial_{\mu}A_{\nu}^{a},
\end{equation} 
it is straightforward to see, that under the divergenceless condition of $\xi^{\mu\nu}_a$, the Lagrangian $L'_{\mathrm{YM}}$ satisfies the same Euler-Lagrange equations as $L_{\mathrm{YM}}$. Moreover, this transformation preserves the De Donder-Weyl Hamiltonian, 
\beqn
H_{\mathrm{DW}}'^{\mathrm{YM}}(A, \pi', x) 
& = &
{\pi'}_a^{\mu\nu} \partial_{\mu}
A^a_{\nu} - L_{\mathrm{YM}}'
\nn \\
& = &\pi_{a}^{\mu\nu}\partial_{\mu}A^{a}_{\nu}-\xi^{\mu\nu}_{a}\partial_{\mu}A_{\nu}^{a}- L_{\mathrm{YM}}+\xi^{\mu\nu}_{a}\partial_{\mu}A_{\nu}^{a} \nn\\
& = & H_{\mathrm{DW}}^{\mathrm{YM}}(A, \pi, x) .
\eeqa
This means that both polymomenta, $\pi^{\mu\nu}_{a}$ 
and ${\pi'}^{\mu\nu}_{a}$, result  physically equivalent. Within this perspective, from the equations (\ref{eq:ym-identities}), we can observe that the symmetric part of the polymomenta satisfies the divergenceless condition, $\partial_{\mu}\pi^{(\mu\nu)}_{a}=0$, this suggest that 
we can define a new set of polymomentum variables given by
\begin{equation}
\pi^{\mu\nu}_a \mapsto \pi'^{\mu\nu}_{a}:= \pi^{\mu\nu}_{a}-\pi^{(\mu\nu)}_{a},
\end{equation} 
such that the equations of motion remain invariant. These new variables correspond to the anti-symmetric part of the polymomenta, therefore, by removing the symmetric part, the weak condition given by the primary constraint (\ref{pcym}), can be taken as a strong condition in the Dirac's sense, thus avoiding the standard formalism for constrained systems. As a consequence of the ambiguity in the De Donder-Weyl equations for the polymomenta, we conclude that the dynamics of the system is fully encoded in its anti-symmetric part as it  is expected.

\subsection{Non-Abelian Chern-Simons model}

In this subsection we will describe, from the polysymplectic point of view, the 
dynamics for the three dimensional Chern-Simons field theory.  
To this end, we start by considering the  Lagrangian~\cite{ferrari,chak} 
\beq
L_{\mathrm{CS}}= \epsilon^{\mu\nu\rho}\left( A^{a}_{\mu} \partial_\nu A_{a\rho} + \frac{1}{3} f_{abc}A^{a}_{\mu}A^{b}_{\nu}A^{c}_{\rho} \right) \,.
\eeq
As before, Greek and lower-case Latin indices stand for space-time and internal symmetry components, respectively.  Also note that 
$A^\mu_a$ represents the gauge fields, $f_{abc}$ are fully anti-symmetric structure constants associated to the gauge algebra and, finally, $\epsilon^{\mu\nu\rho}$ stands for the standard three dimensional Levi-Civita symbol. Proceeding as in the previous subsection, we start by considering the definition \eqref{poly} in order to obtain the polymomenta
\beq\label{eq:csmomentum}
\pi^{\mu\nu}_{a} 
= 
\frac{\partial L_{\mathrm{CS}}}{\partial(\partial_\mu A^a_\nu)}
= 
\epsilon^{\mu\nu\rho} A_{a\rho} \,.
\eeq
From this last relation we note that, contrary
to the Yang-Mills case, all the derivatives 
of the gauge fields are not invertible in 
terms of the polymomenta.  Thus, we obtain
a set of primary constraints that we separate
for convenience into symmetric and anti-symmetric parts
\begin{eqnarray}
\label{cspc}
\pi^{(\mu\nu)}_{a} 
& \approx & 
0 \,, 
\nn\\
\pi^{[\mu\nu]}_{a} -\epsilon^{\mu\nu\rho} A^{a}_{\rho} 
& \approx & 
0 \,. 
\end{eqnarray}
Once again, by means of the covariant Legendre transformation~\eqref{legen} we compute the De Donder-Weyl Hamiltonian associated to the Chern-Simons model
\beqn
H_{\mathrm{DW}}^{\mathrm{CS}}(A, x) 
& = &  
\pi_a^{\mu\nu} \partial_{[\mu}
A^a_{\nu]} - L_{\mathrm{CS}}
\nn \\
& = &
-\frac{1}{3} \epsilon^{\rho\mu\nu}f_{abc}A^{a}_{\rho}A^{b}_{\mu}A^{c}_{\nu} \,,
\eeqa
and, in analogy to the Yang-Mills case, we may introduce the total Hamiltonian 
\begin{equation}\label{CSTH}
\widetilde{H}_{\mathrm{DW}}^{\mathrm{CS}}  = H_{\mathrm{DW}}^{\mathrm{CS}} + \lambda^{a}_{\mu\nu}\pi^{(\mu\nu)}_{a} + \eta^{a}_{\mu\nu}(\pi^{[\mu\nu]}_{a}-\epsilon^{\mu\nu\rho} A^{a}_{\rho}) \,,
\end{equation}
where the $\lambda$'s and the $\eta$'s are, respectively, symmetric and anti-symmetric  Lagrange multipliers enforcing the 
constraints~\eqref{cspc}.   As expected, 
we will use this total Hamiltonian in order to reproduce the correct equations of motion 
as defined in~\eqref{dwfe} by means of the 
Poisson-Gerstenhaber brackets.  
To this end, we define the canonical $(n-1)$-form  variables $A^{a\nu}_\mu:=A^a_\mu 
\omega^\nu$ and $\pi_a^\nu:=\pi^{\mu\nu}_a \omega_\mu$. Thus, the field equations read
\beqn
\label{csaaa}
\textbf{d} \bullet A^{a}_{\mu\nu} 
& = &  
- \pgb \widetilde{H}_{\mathrm{DW}}^{\mathrm{CS}}, A_{\mu\nu}^a\pbg  %+ d^h \bullet A_{\mu\nu}^a
\nn\\
& = &
\lambda^{a}_{\mu\nu}+\eta^{a}_{\mu\nu} \,,
\nn\\
\textbf{d} \bullet \pi_{a}^{\mu} 
& = &
- \pgb \widetilde{H}_{\mathrm{DW}}^{\mathrm{CS}}, \pi_a^\mu\pbg  %+ d^h \bullet \pi_a^\mu  
\nn\\ 
& = & 
\epsilon^{\mu\nu\rho}f_{abc}A^{b}_{\nu}A^{c}_{\rho}+\eta_{a\nu\rho}\epsilon^{\mu\nu\rho} \,,
\eeqa
Once again, one may consider the 
symmetric and anti-symmetric parts of these field equations, thus obtaining  the 
relations
\beqn
\label{eq:cs-identities}
\partial_{[\mu}A^{a}_{\nu]} 
& = & 
\eta^{a}_{\mu\nu} \,,
\nn\\ 
\partial_{(\mu}A^{a}_{\nu)} 
& = &  
\lambda^{a}_{\mu\nu} \,, 
\nn\\
\partial_{\mu}\pi^{[\mu\nu]}_{a} 
& = & \epsilon^{\nu\mu\rho}f_{abc}A^{b}_{\mu}A^{c}_{\rho} 
-\eta_{a\mu\rho}\epsilon^{\nu\mu\rho} \,,
\nn\\
\partial_{\nu}\pi^{(\nu\mu)}_{a} 
& = & 
0 \,.
\eeqa
This set of relations may be interpreted 
as follows.  The first two lines simply 
fix both Lagrange multipliers, $\eta^a_{\mu\nu}$ and 
$\lambda^a_{\mu\nu}$, in terms of 
the gradients of the gauge field.  
Likewise, in order to obtain the Chern-Simons field equations we only need to substitute,
respectively, 
the second line of the constraints~\eqref{cspc} on the left hand side of the third equation 
and also the first line of equations~\eqref{eq:cs-identities} in the 
same line to obtain  in a direct manner
\begin{eqnarray}
\label{eq:cs-eom}
\epsilon^{\alpha\mu\nu}F_{\mu\nu}^a = 0 \,,\end{eqnarray}
where we defined the components of the Chern-Simons 3-form $F_{\mu\nu}^a$ as in~\eqref{fcomp}. \cite{chak} 
Of course, equations~\eqref{eq:cs-eom} may be recognized as the De Donder-Weyl field equations of the non-Abelian Chern-Simons field theory. Finally, in a similar fashion as for the Yang-Mills model, the first line of constraints~\eqref{cspc}, together with the last line of relations~\eqref{eq:cs-identities} tell us that the symmetric part of the polymomentum does not contain any dynamical information about the Chern-Simons
model.  
As in the previous example, we can observe from the last equation in~\eqref{eq:cs-identities}, that the polymomenta satisfy the divergenceless condition $\partial_{\nu}\pi^{(\nu\mu)}_{a}=0$. This means that the transformed polymomentum variables
$
\pi^{\mu\nu}_a \mapsto \pi'^{\mu\nu}_{a} := \pi^{\mu\nu}_{a}-\pi^{(\mu\nu)}_{a}
$
leave the De Donder-Weyl equations invariant, therefore, the dynamics of the system is fully determined by its anti-symmetric part. Since the symmetric part of the polymomenta $\pi^{(\mu\nu)}_{a}$ is related to the primary constraints, its divergenceless condition suggests that these constraints can be taken as strong constraints in the Dirac sense.

\subsection{Topologically massive Yang-Mills theory}

Based on the two previous examples, we will consider now the Lagrangian for the three dimensional topologically massive Yang-Mills field theory given by~\cite{tmym} 
\begin{equation}
L_{\mathrm{TMYM}} = -\frac{1}{4} F_{\mu\nu}^a F^{\mu\nu}_a + \frac{m}{4}\epsilon^{\mu\nu\rho}\left( F_{a\mu\nu} A^a_\rho - \frac{g}{3} f_{abc} A^a_\mu A^b_\nu A^c_\rho \right) \,,
\end{equation}
where $m$ and $g$ are free parameters commonly associated to the mass of the field and to the coupling constant, respectively, 
while $A^a_\mu$ stands 
for the gauge field. The components of 
the field strength $F_{\mu\nu}^a$ are 
analogously given by 
equation~\eqref{fcomp}, and $f_{abc}$ are 
again the structure constants associated to 
the gauge symmetry of this model.
In order to construct the De Donder-Weyl 
Hamiltonian first we will introduce the  polymomenta by means of~\eqref{poly}
\beqn
\label{tmymmom}
\pi^{\mu\nu}_a 
= 
\frac{\partial L_{\mathrm{TMYM}}}{\partial(\partial_\mu A^a_\nu)}
=
-F^{\mu\nu}_a + \frac{m}{2} \epsilon^{\mu\nu\rho} A_{a\rho} \,,
\eeqa
which systematically may be decomposed into 
their symmetric and anti-symmetric parts as
\beqn
\label{eq:tmym-momenta}
\pi^{(\mu\nu)}_a 
& \approx & 
0 \,, 
\nn\\
\pi^{[\mu\nu]}_a
& = & -2\partial^{[\mu} A^{\nu]}_a
-g f^{\ bc}_a A^\mu_b A^\nu_c \pi^{[\mu\nu]}_{a} + \frac{m}{2}\epsilon^{\mu\nu\rho}A_{a\rho}
\,.
\eeqa
Note that only the symmetric part of the polymomenta $\pi^{\mu\nu}_a$ may be identified with a primary constraint as, from the last line above, we may clearly see that the 
derivatives of the gauge field $A_\mu^a$ are invertible in terms of the anti-symmetric components of the polymomenta.  Next, we adapt
Legendre transformation~\eqref{legen} to our 
system encountering the De Donder-Weyl Hamiltonian for the topologically massive 
Yang-Mills field
\begin{eqnarray}
H_{\mathrm{DW}}^{\mathrm{TMYM}}(A, \pi, x) 
& = & 
\pi_a^{[\mu\nu]} \partial_{[\mu}
A^a_{\nu]} - L_{\mathrm{TMYM}}
\nn \\ 
& = &  
-\frac{1}{4} \pi_{[\mu\nu]}^a \pi^{[\mu\nu]}_a 
-\frac{g}{2} f^{\ bc}_a A^\mu_b A^\nu_c \pi_{[\mu\nu]}^a  
+ \frac{m}{4}\epsilon^{\mu\nu\rho}A_{a\rho} \pi_{[\mu\nu]}^a 
\nn \\  
&  & 
-\frac{m^2}{16}\epsilon^{\mu\nu\rho}\epsilon_{\mu\nu\rho} A_a^\rho  A^a_\rho 
+ \frac{mg}{12}\epsilon^{\mu\nu\rho}f_{abc}A^a_\mu A^b_\nu A^c_\rho \,.
\end{eqnarray}
Once again, by taking into account the constraints for the symmetric part of the 
polymomenta, we propose the total Hamiltonian
\begin{eqnarray}\label{tmtmtotham}
\widetilde{H}_{\mathrm{DW}}^{\mathrm{TMYM}} = 
H_{\mathrm{DW}}^{\mathrm{TMYM}} + \lambda^a_{\mu\nu}\pi^{(\mu\nu)}_a \,,
\end{eqnarray}
where the $\lambda^a_{\mu\nu}$ stand for 
Lagrange multipliers enforcing the constraints, as it is common by now.
The De Donder-Weyl equations 
are obtained from~\eqref{dwfe} when applied 
to the canonical pair of $(n-1)$-forms 
$(A_a^{\mu\nu},\pi_a^\mu)$ defined as in the previous examples, thus yielding
\beqn
\label{tmymfaaa}
\textbf{d}\bullet A^{\mu\nu}_a 
& = & 
- \pgb \widetilde{H}_{\mathrm{DW}}^{\mathrm{TMYM}}, A^{\mu\nu}_a\pbg  % + d^h \bullet A^{\mu\nu}_a  
\nn\\
& = & 
-\frac{1}{2} \pi^{[\mu\nu]}_a -\frac{g}{2} f^{\ bc}_a A^\mu_b A^\nu_c  + \frac{m}{4}\epsilon^{\mu\nu\rho}A_{a\rho} + \lambda_a^{\mu\nu} 
\,,
\nn\\
\textbf{d}\bullet \pi_a^\mu 
& = &
- \pgb \widetilde{H}_{\mathrm{DW}}^{\mathrm{TMYM}}, \pi_a^\mu\pbg  % + d^h \bullet \pi_a^\mu
\nn\\
& = & 
gf_{a}^{\ bc} \pi^{[\mu \nu]}_b A_{c\nu} 
- \frac{m}{4}\epsilon^{\mu\nu\rho}\pi_{a[\nu\rho]} 
-\frac{mg}{4}f_{abc}\epsilon^{\mu\nu\rho} A^b_{[\nu} A^c_{\rho]} \nonumber \\
&   & 
+ \frac{3m^2}{4}  A^\mu_a \,.
\eeqa
Proceeding as in the previous subsections, for the sake of simplicity we decomposed the above equations into their symmetric and 
anti-symmetric parts, obtaining the identities
\begin{eqnarray}
\label{tmymiden}
\partial^{[\mu} A^{\nu]}_a 
& = & 
-\frac{1}{2} \pi^{[\mu\nu]}_a -\frac{g}{2} f^{\ bc}_a A^\mu_b A^\nu_c  + \frac{m}{4}\epsilon^{\mu\nu\rho}A_{a\rho} \,,
\nn\\
\partial^{(\mu} A^{\nu)}_a 
& = & 
\lambda_a^{(\mu\nu)} \,, 
\nn\\
%\label{tmymfe}
\partial_\nu \pi^{[\nu\mu]}_a
& = & 
gf_{a}^{\ bc} \pi^{[\mu \nu]}_b A_{c\nu} - \frac{m}{4}\epsilon^{\mu\nu\rho}\pi_{a[\nu\rho]} -\frac{mg}{4}f_{abc}\epsilon^{\mu\nu\rho} A^b_{[\nu} A^c_{\rho]} \nn\\
&   & 
+ \frac{3m^2}{4}  A^\mu_a \,,
\nn\\
\partial_\nu \pi^{(\nu\mu)}_a 
& = & 
0 \,.
\end{eqnarray}
Interpretation of the above identities
follows in a similar manner as in 
the previous examples.   
The first line stands for the invertibility of the derivatives of the field in terms of the polymomenta and the gauge fields.  
The second line fixes the Lagrange multipliers 
$\lambda_a^{(\mu\nu)}$.  Further, by considering the 
second line of~\eqref{eq:tmym-momenta} and the definition 
of the field strength $F^{\mu\nu}_a$ adapted to
this model in analogy to~\eqref{fcomp}, we straightforwardly find 
the De Donder-Weyl field equations for the
topologically massive Yang-Mills theory 
\begin{eqnarray}
\label{tmymfe}
D^b_{\nu a} F^{\mu\nu}_b -\frac{m}{2}\epsilon^{\mu}_{\ \nu\rho} F^{\nu\rho}_a 
&=& 0 \,,
\end{eqnarray}
where $D^{a}_{\nu b}$ is a covariant derivative defined in analogy to~\eqref{eq:covderivative}.
Finally, as in the preceding two examples we expect that only the anti-symmetric part of the polymomentum is truly associated to the dynamics of the model. In order to see this, by using the last equation appearing in (\ref{tmymiden}), we define a new set of polymomentum variables by
$
\pi^{\mu\nu}_{a} \mapsto \pi'^{\mu\nu}_{a} := \pi^{\mu\nu}_{a}-\pi^{(\mu\nu)}_{a}
$.
The divergenceless condition  of the polymomenta $\partial_{\nu}\pi^{(\nu\mu)}_{a}=0$, means that its symmetric part can be removed from the De Donder-Weyl equations of motion without modifying the dynamics of the system. In conclusion, due to the symmetric properties of this model we note that within our 
formulation we may avoid the standard Dirac procedure 
for constrained systems in order to obtain the 
correct De Donder-Weyl equations of motion which turn out to be totally equivalent to 
the Lagrangian field equations.

\section{Conclusions}
\label{conclusion}

We analyzed several non-Abelian field theoretical models within the polysymplectic framework.  In particular, we 
consider the topologically massive Yang-Mills field theory which is described by adding a Chern-Simons invariant to the Yang-Mills theory.  
In this way, we started by introducing the polysymplectic 
formalism  for the Yang-Mills and Chern-Simons theories
separately, and then we proceed systematically.
Indeed, all of these theories present similar 
characteristics as all of them are described by 
a singular Lagrangian  which, however, allowed us to 
find proper polymomenta in such a way that we were able to avoid 
the constraint analysis by introducing 
a completely equivalent De Donder-Weyl Hamiltonian.  For any of the theories 
considered, their respective equivalent Hamiltonians 
may be inherited from the De Donder-Weyl formalism due to the presence of divergenceless of the symmetric part 
of the polymomenta.  Further, 
for these new 
equivalent De Donder-Weyl Hamiltonians it was completely clear that
only the anti-symmetric part of the 
polymomenta resulted relevant, as expected.  

Besides, an important issue allowed by our 
separation of the 
canonical variables $(n-1)$-forms 
into their symmetric and anti-symmetric parts was to keep our description of the 
field equations in a closed way from the perspective of 
the  Poisson-Gerstenhaber bracket, 
thus strengthen the primary role that the $(n-1)$
forms play within the polysymplectic formalism.  
This must be confronted with several other 
examples found in the literature (see for 
example,~\cite{kan} and~\cite{helein1}) for which one may, in contrast, incorporate other than $(n-1)$-forms as 
canonical variables in order to get the correct 
equations of motion.

We hope that the introduced polysymplectic formalism
for the topologically massive field theory may 
shed some light on the quantization for these kind 
of theories from the polysymplectic point of view.  
In particular, it will be relevant 
to test Kanatchikov's Schr\"odinger-like 
quantum equation (see~\cite{kan2} for further details) as, in principle, one may wonder in which sense this quantum scheme may incorporate 
the cases for non-regular Lagrangian theories.
Even though there are recent efforts in this 
direction (see for example~\cite{kanGR1} and~\cite{kanGR2} in the context of Einstein gravity),
we think that the non-Abelian field theories 
analyzed above may serve as a starting point 
for the  analysis of the quantum version of non-regular Lagrangians from the polysymplectic perspective.
This will be done elsewhere.

\section*{Acknowledgments}
The authors would like to thank 
Jos\'e A.~Zapata 
and Jairo Mart\'inez-Montoya for fruitful discussions 
on the multisymplectic formalism and David Serrano-Blanco for collaboration
and discussion.  AM acknowledges financial
support from CONACYT-Mexico under project CB-2014-243433.

% \section*{REFERENCES}

\end{document}